\documentclass{PoS}
\usepackage{mathrsfs}
\usepackage{psfrag}
\PoS{PoS(LAT2005)320}
\title{Center Vortex Model for SU(3) Yang-Mills Theory }
\ShortTitle{Center Vortex Model for SU(3) Yang-Mills Theory }
\author{\speaker{Markus Quandt}, Hugo Reinhardt\\
        Institut f\"ur Theoretische Physik \\
	Universit\"at T\"ubingen\\
        E-mail: \email{quandt@tphys.physik.uni-tuebingen.de}}
\author{Michael Engelhardt\\
	Physics Department\\
	New Mexico State University, Las Cruces\\
	E-Mail: \email{engel@nmsu.edu}}

\abstract{The center vortex model for the infrared sector of $SU(3)$ Yang-Mills theory
is reviewed. After discussing the physical foundations underlying the model,
some technical aspects of its realisation are discussed. The confining properties
of the model are presented in some detail and compared to known
results from full lattice Yang-Mills theory. Particular emphasis is put on
the new phenomenon of vortex branching, which is instrumental in establishing
first order behaviour of the $SU(3)$ phase transition. Finally, the vortex
free energy is verified to furnish an order parameter for the
deconfinement phase transition. It is shown to exhibit a weak discontinuity
at the critical temperature, in agreement with predictions from
lattice gauge theory.}

\FullConference{XXIIIrd International Symposium on Lattice Field Theory\\
		 25-30 July 2005\\
		 Trinity College, Dublin, Ireland}

\begin{document}

\section{Introduction}

The vortex picture of the Yang-Mills vacuum was initially proposed as a
possible mechanism of colour confinement. It is based on the idea that a
random distribution of vortex colour flux is sufficient to effect an area law
behaviour for large Wilson loops. Despite its simplicity and early
successes \cite{V1}, a precise confirmation of this scenario has been elusive for a
long time. Recently, the picture has attracted renewed attention, mainly due
to the advent of gauge fixing techniques which allow to detect center vortex
structures directly within lattice configurations. Large-scale computer
simulations have revealed ample evidence that the long-range properties of
Yang-Mills theory can be accounted for in terms of vortices \cite{V3}.

To complement the lattice approach, a random vortex world-surface model was
introduced as an effective low-energy description of $SU(2)$ Yang-Mills
theory \cite{V4}; it has recently been extended to the gauge groups $SU(3)$
\cite{V5,V6}. The fundamental assumption is that the long-range
structure of Yang-Mills theory is dominated by thick, weakly interacting
tubes of center flux which trace out closed surfaces in space-time.
No analytical approach is known for such a quantum string ensemble. As a
consequence, we realise our model on a space-time lattice with a \emph{fixed}
spacing $a$ to represent the transverse thickness of vortices. Short-distance
structures smaller than $a$ (or, equivalently, momenta larger than $\pi/a$)
cannot be resolved in this approach.

I will discuss the physical foundations of the center vortex model
and some technical aspects in the next section. The main part of the
talk is section 3 which presents a selection of results obtained in the
model. In Section 4, I conclude with a brief summary and some
general comments.

\section{Phase diagram and choice of parameters}

Vortices are closed lines of colour flux in three space dimensions and,
correspondingly, closed world-surfaces in space-time.
Their dimensionality is unique in that they can have a topologically stable
linking with Wilson loops. The colour flux placed on the vortex is quantized
such that each linking with a Wilson loop $W(\mathscr{C})$ (i.e. each
\emph{intersection point} of the closed loop $\mathscr{C}$ with the vortex surface)
contributes a center element of $SU(N)$ to $\langle W(\mathscr{C}) \rangle$.
To describe such intersections, the vortex surface must be
defined on a lattice which is \emph{dual} to the one where Wilson loops
(and gauge connections) live. This emphasises the dual character of our model.

Technically, we create random vortex surfaces on the dual lattice by assigning
\emph{triality}\footnote{For the case $G=SU(3)$, the center
comprises three elements $Z(3) = \{ 1,\, e^{2\pi i/3},\, e^{4 \pi i/3} \}$.
They are usually parametrised as $z=\exp(2\pi i /3 \cdot q)$ in terms of the
\emph{triality} $q \in \{0,1,2 \}$ defined modulo 3. The triality of an elementary
square determines the center element which a Wilson loop receives when intersecting
the square.}
to the elementary squares. The fixed lattice spacing $a$ defines
the minimal distance at which two intersection points can be resolved, i.e. the
thickness of the vortex tube or sheet. Random surfaces created in this way are
weighted by a model action containing a Nambu-Goto and a curvature term \cite{V5},
symbolically
\begin{equation}
S = \sum_{\rm plaq} \epsilon\,\,\,
\begin{minipage}{15mm}\includegraphics[width=15mm]{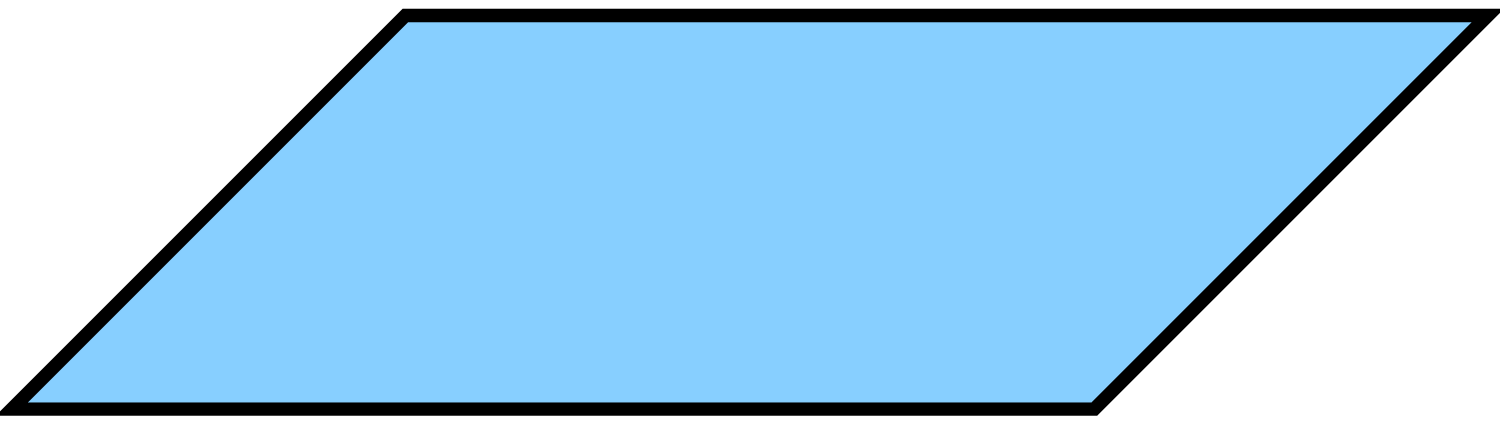}
\end{minipage} +
\frac{c}{2} \quad\begin{minipage}{15mm}
\includegraphics[width=15mm]{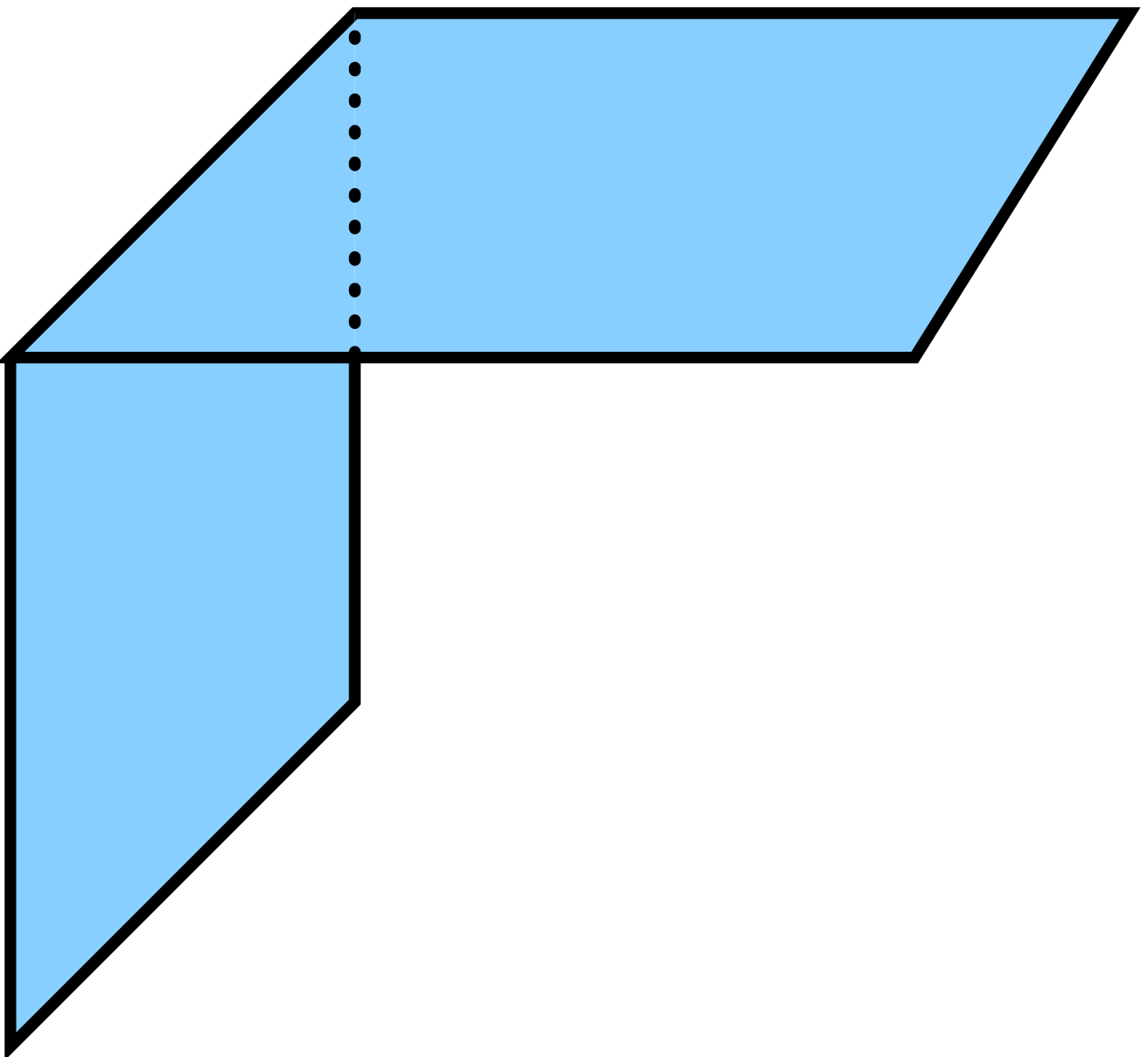} \end{minipage}
\label{eqn1}
\end{equation}
The parameters are determined by measuring the zero-temperature string tension
$\sigma_0$ (in units of the lattice spacing). One observes a deconfined region
at large couplings $(\epsilon,c)$ with a shallow cross-over to a confined
region at small $(\epsilon,c)$ \cite{V5}. In the latter domain,
one can also effect a deconfinement phase transition by increasing the temperature
of the simulation (i.e. by reducing the temporal lattice extension). Using an
interpolation procedure, the ratio of the critical temperature,
$T_c$, to the zero-temperature string tension, $\sigma_0$, can be extracted at
all couplings. Comparision with the known value $T_c / \sqrt{\sigma_0} = 0.63$
from $SU(3)$ lattice gauge theory then yields a \emph{line} of possible physical values
in parameter $(\epsilon,c)$ space. As it turns out, the long-range physics are almost
constant on this parameter line so that we are free to make the arbitrary choice,
\begin{equation}
\epsilon = 0\,,\qquad c = 0.21\,.
\label{eqn2}
\end{equation}
From the string tension $\sigma_0\,a^2$ measured at this point,
the overall scale can be determined using the phenomenological value
$\sigma_0 = (440 \,\mathrm{MeV})^2$ from QCD. This gives a vortex thickness of
$a = 0.39\,\mathrm{fm}$.


\section{Applications}

\begin{figure}[t]
\centerline{
\begin{minipage}{7cm}
\includegraphics[width = 7cm]{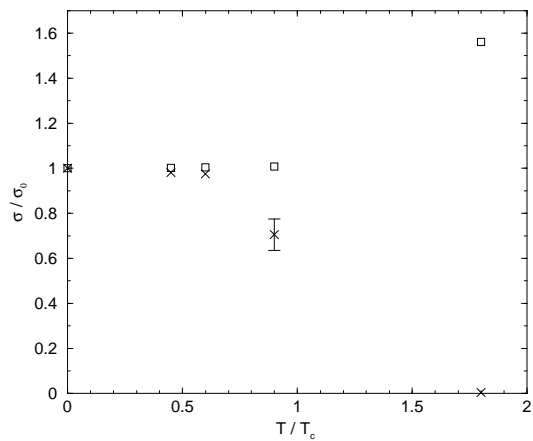} \end{minipage}
}
\caption{String tension between two static colour charges
(\emph{crosses}) and spatial string tension (\emph{squares}) as a
function of temperature. Measurements were taken on a $16^3 \times N_0 $
lattice for the physical choice of parameters eq.~(2.2).}
\label{fig2}
\end{figure}

\subsection{Spatial string tension}

The zero-temperature string-tension $\sigma_0$ used to fix the parameters
is extracted from Wilson loops extending in one space and the time direction.
In addition, purely spatial Wilson loops can also be measured, although they
do not have an immediate interpretation as an inter-quark potential.
From lattice gauge theory, it is known that spatial Wilson loops exhibit an
area law at \emph{all} temperatures and the \emph{spatial string tension}
extracted from them persists in the high-temperature phase \cite{V10}.
As shown in fig.~\ref{fig2}, our model reproduces this strongly
correlated hot phase; in fact, our results for the spatial string tension
agree with ref.~\cite{V10} to within $5\%$.
This success can be attributed to the \emph{percolation} behaviour of
vortex clusters.

\subsection{Finite temperature phase transition and vortex branching}
\label{branching}

The confinement mechanism is most clearly seen in the details of the phase
transition at non-zero temperatures.
Histograms of the action density measured at the critical temperature
reveal a qualitative difference between the gauge groups $SU(2)$ and $SU(3)$:
The $SU(3)$ transition exhibits the shallow double-peak structure characteristic
for a weak \emph{first order transition}, while the $SU(2)$ transition is
continuous (\emph{second order}) \cite{V5}. This pattern is also seen
in full lattice gauge theory.

\begin{figure}[t]
\centerline{
\hfill
\begin{minipage}{4cm}
\includegraphics[width = 4cm]{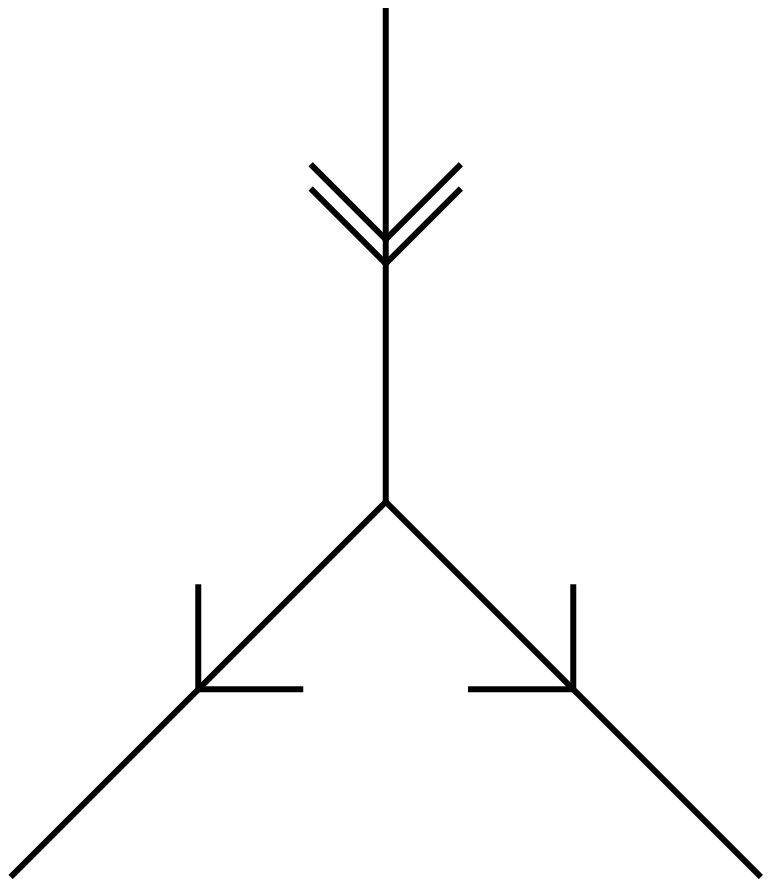} \end{minipage}\hfill
\begin{minipage}{4cm}
\includegraphics[width = 4cm]{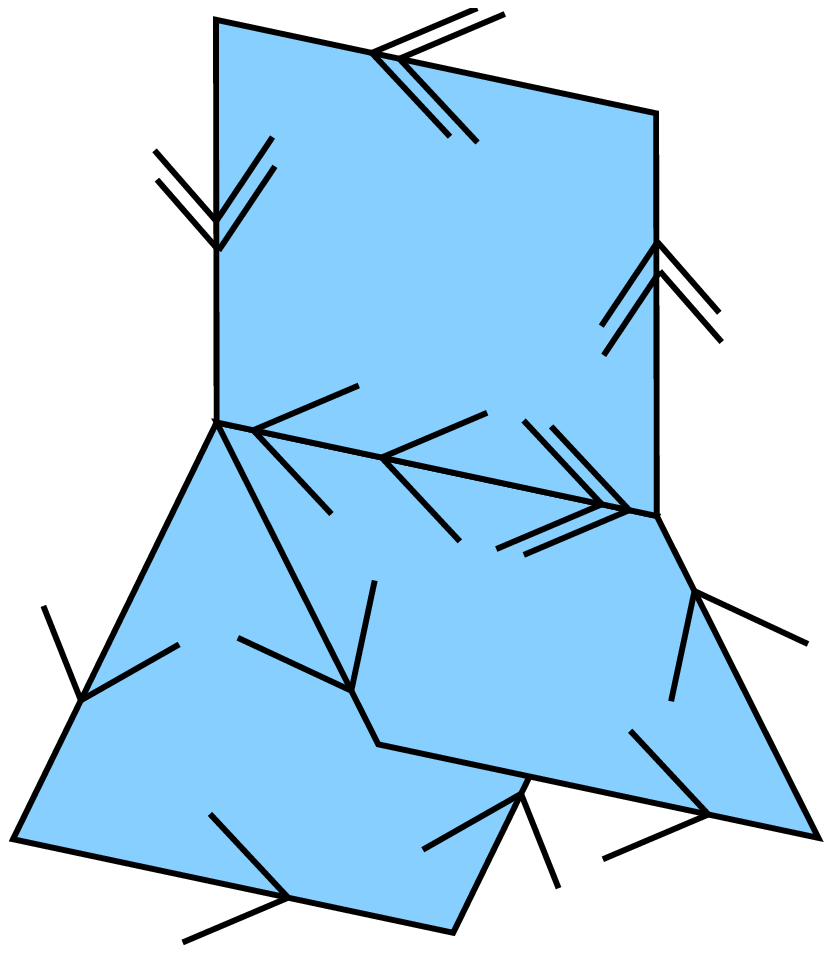}\end{minipage}
\hfill
}
\caption{The right-hand panel displays a four-dimensional lattice view,
where a vortex surface, consisting of elementary squares, branches along
a $\nu=3$ bond. The left panel shows the same situation projected to a 3D slice.}
\label{fig4a}
\end{figure}

To understand the origin of this difference, we have to look at
the vortex geometry in more detail. In $D=4$ dimensions, each bond
on the dual lattice is attached to six elementary squares, which are
assigned trialities $q \in \{0,1,2\}$ in our model.
The local geometry of a world surface is then characterised by the number
$\nu = 0,\ldots,6$ of vortex surfaces ($q\neq 0$) meeting at a given bond.
The odd values $\nu=3,5$ are not allowed in $SU(2)$ and represent genuine
$SU(3)$ \emph{vortex branchings}, cf.~fig.~\ref{fig4a}. Note that the
triality $q=1,2$ of a vortex square can be reversed by flipping its
orientation; moreover, triality is only conserved modulo 3 and a
Dirac string ($q=3$) may be added to any configuration.
The branching in fig.~\ref{fig4a} is thus equivalent to a situation
where three $q=1$ vortices emerge from a common point (or bond in $D=4$).
Branching lines in $D=4$ are therefore similar to \emph{center monopole}
worldlines.

\begin{figure}
\hfill
\includegraphics[height=.17\textheight]{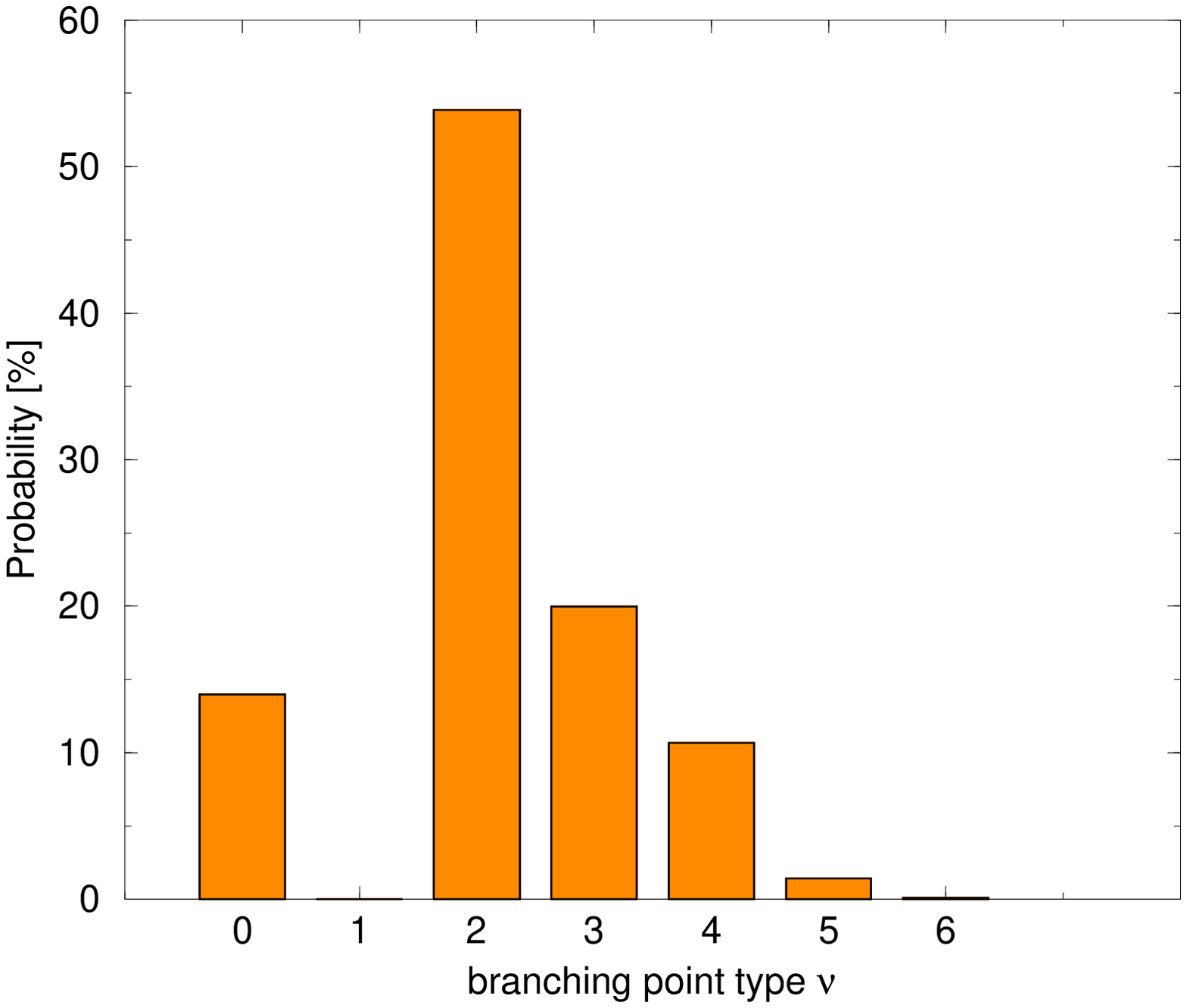}
\hspace{.5cm}
\includegraphics[height=.17\textheight]{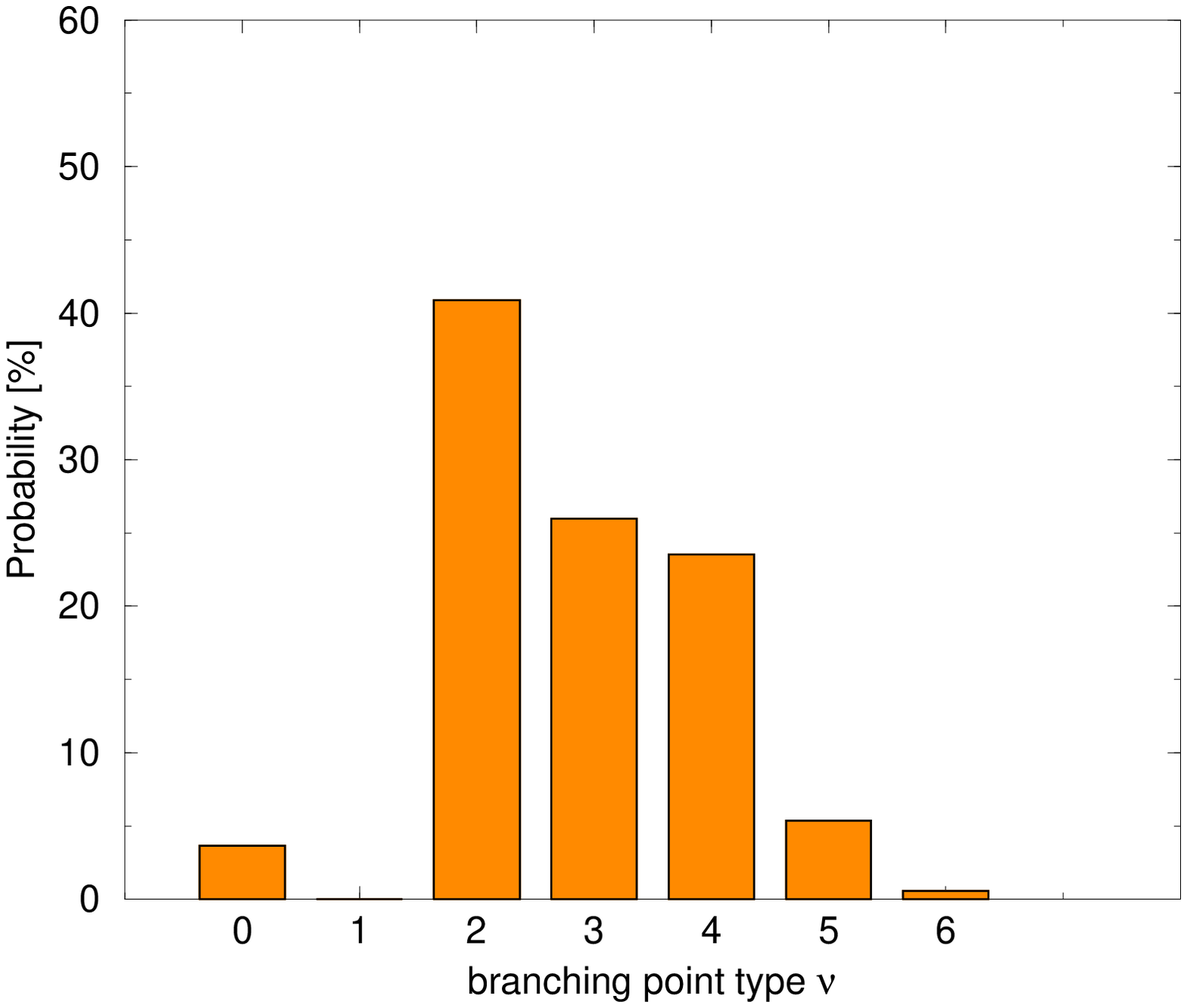}
\hspace{.5cm}
\includegraphics[height=.17\textheight]{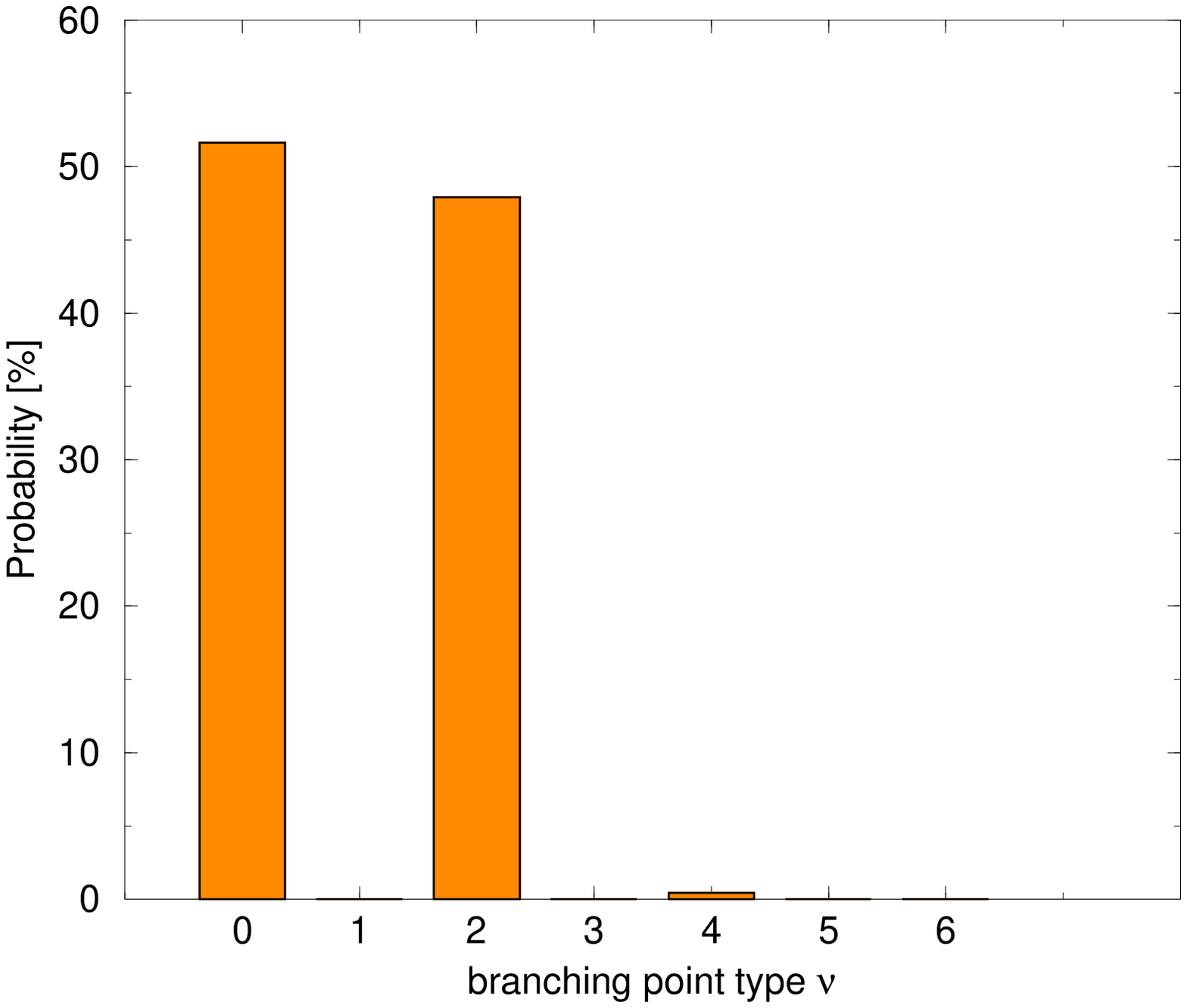}
\hfill
\caption{Volume fractions occupied by points of a certain branching type
$\nu$ within $3D$ lattice slices. The left panel shows the distribution at
zero temperature in the confined phase. The middle and right panel
both correspond to $T>T_c$, with the middle referring to a \emph{time slice}
and the right to a \emph{space slice}.}
\label{fig4}
\end{figure}

The statistical distribution of branchings is best studied in
3D slices of the lattice, whence possible branching bonds are projected onto
\emph{points} of type $\nu$. From fig.~\ref{fig4}, we conclude that the largest
volume fraction in the confined phase corresponds to non-branching vortex
matter ($\nu=2$), with a considerable probabitlity of both
self-intersection ($\nu=4,6$) and genuine branchings ($\nu=3,5$). Only $15 \%$
of the volume is not occupied by vortices ($\nu=0$).\footnote{The
case $\nu=1$ describes a single vortex surface \emph{ending} in the given bond;
this is forbidden by Bianchi's identity.} In the deconfined phase ($T > T_c$),
the situation is qualitatively unchanged for \emph{time-slices}, while
\emph{space-slices} show virtually zero branchings above $T_c$. This can be understood
if the vortices undergo a \emph{(de)percolation phase transition} for
$T > T_c$ and most vortex clusters wind directly around the short time
direction \cite{V5}. The same phenomenon explains the persistence of the
spatial string tension discussed above.

\subsection{Vortex free energy and the 't Hooft loop}

The 't Hooft loop  can be viewed as a creation operator
for (quantized) flux along a closed spatial line $\mathscr{C}$. It was
formally introduced by 't Hooft in 1978 as a dual order parameter
for the deconfinement transition on a space-time torus. Recently, explicit
realisations of this construction have been given both in the continuum
\cite{V8} and on the lattice \cite{V9}, where maximally extended
't~Hooft loops implement twisted boundary conditions.

In the center vortex model, the 't Hooft loop is literally an
(open) \emph{vortex creation operator} \cite{V6}: Its action is to add a
fixed triality $q=1,2$ to each elementary square in a world-sheet over
$\mathscr{C}$; since triality is additive, this simply injects a center
vortex of type $q=1,2$ in the current configuration. The exact form of the
world-sheet over $\mathscr{C}$ is irrelevant (and center-gauge dependent),
but for simplicity, we restrict ourselves to planar loops $\mathscr{C}$
and minimal surface sheets over them.

\begin{figure}
\centerline{
\includegraphics[height=.215\textheight]{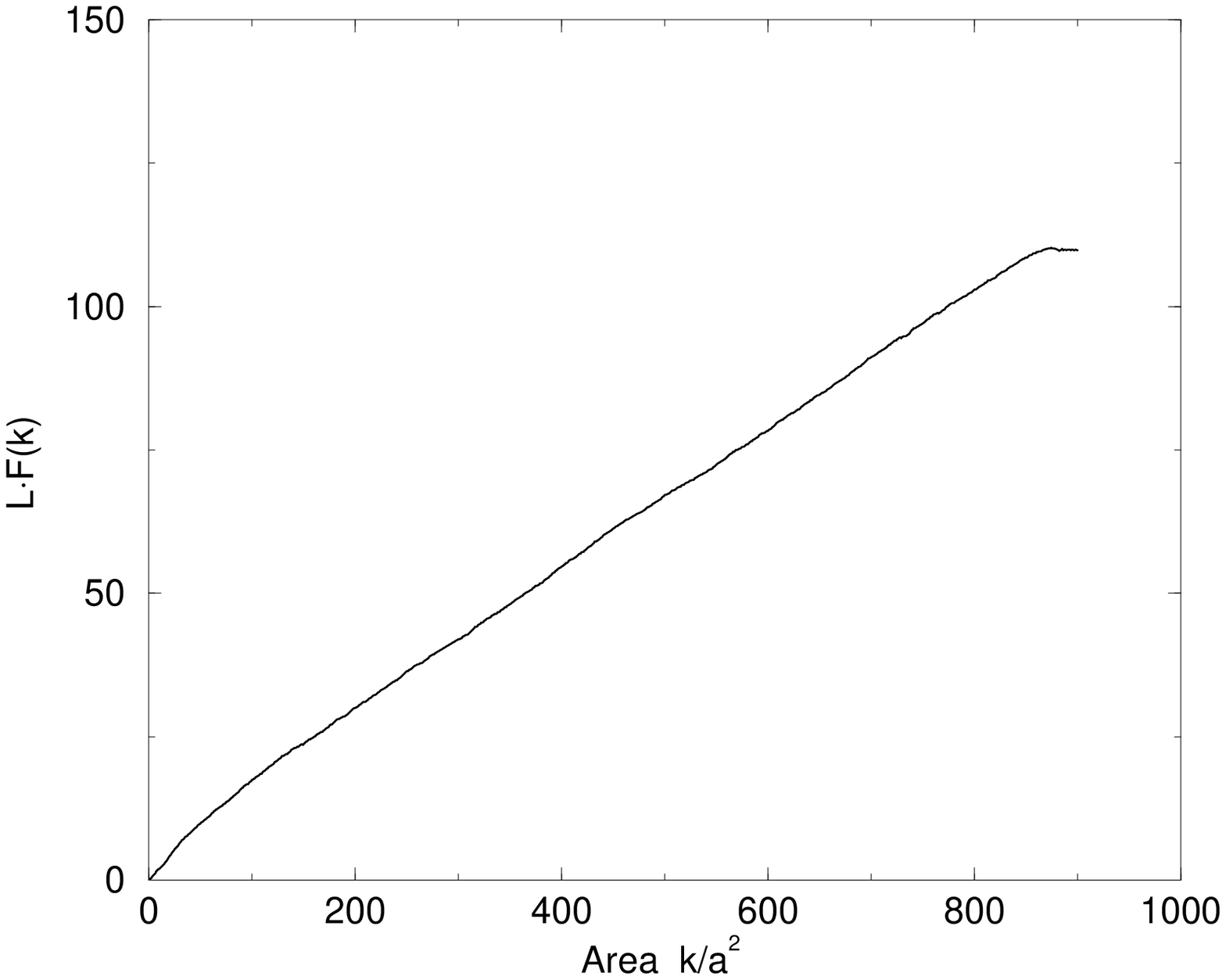}
\hspace{1cm}
\includegraphics[height=.215\textheight]{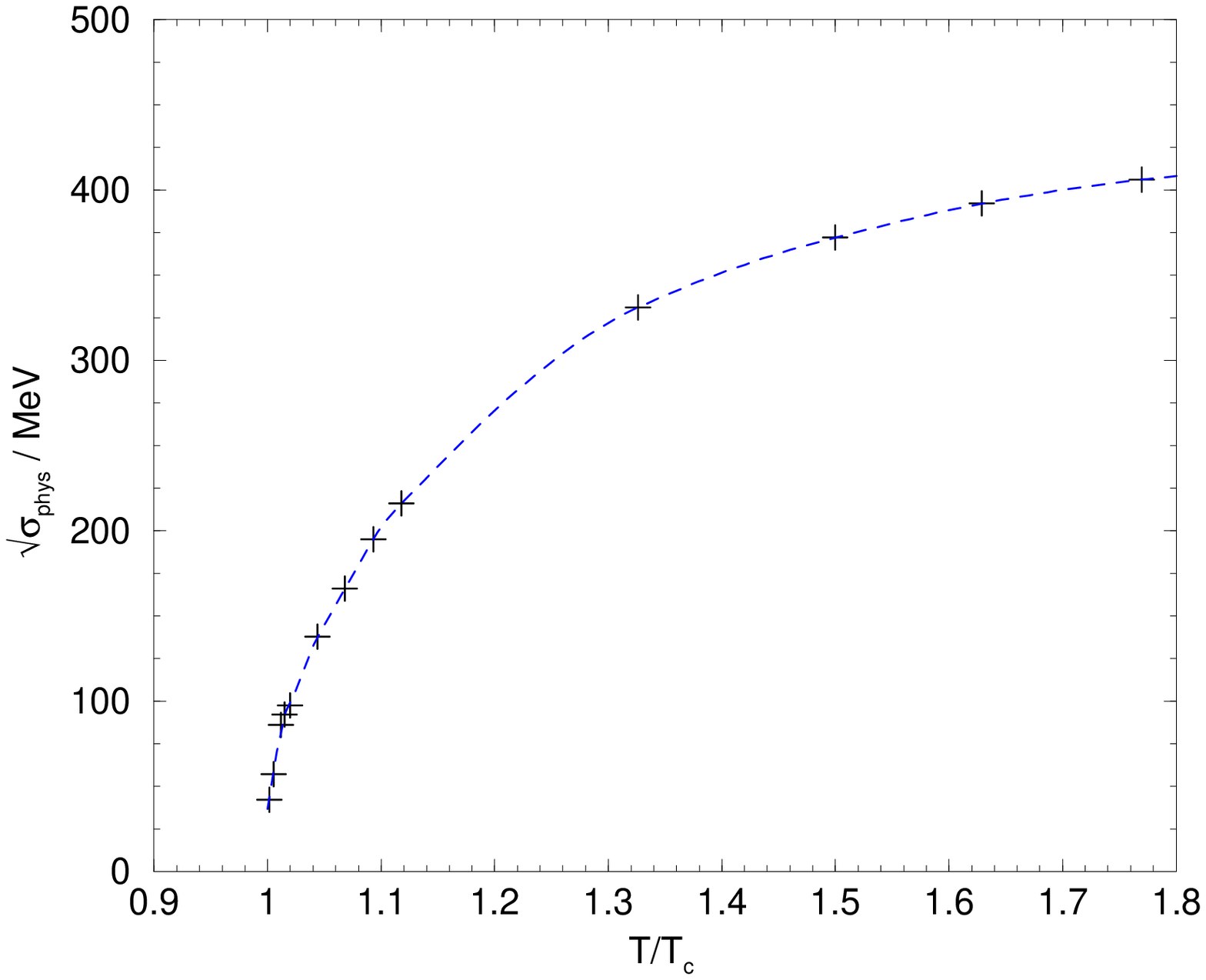}
}
\caption{Left Panel: Free energy of 't Hooft loops in the
deconfined phase ($T/T_c = 1.093$), as a function of the minimal area over
the loop. Right panel: The dual string tension $\sigma_D$ as a function of
the temperature. Measurements were performed on a large
$30^3 \times N_0$ lattice with $N_0 = 1,2$.}
\label{fig5}
\end{figure}

The action penalty $\Delta S$ incurred by the vortex creation is related to
the \emph{vortex free energy} $F$ via $e^{-F} = \langle e^{-\Delta S} \rangle$.
As mentioned above, the free energy is expected to
furnish an order parameter for the deconfinement transition with
essentially the opposite behaviour as the Wilson loop. This is nicely
confirmed in our model:
The left panel of fig.~\ref{fig5}
exhibits a linear rise of the free energy with the area of the 't Hooft loop
at $T > T_c$,\footnote{The systematic deviations can be understood in terms of
$Z(3)$ monopole correlations, see \cite{V6} for details} which allows to
define a \emph{dual string tension}  $\sigma_D$ in the deconfined phase.

As we approach the phase transition from above, the dual string tension
quickly vanishes (cf.~right panel of fig.~\ref{fig5}). In the confined
phase, the subleading perimeter law is hidden in the statistical noise,
and the vortex free energy is consistent with zero (\emph{vortex condensation}).
Precise measurements close to the transition reveal a small discontinuity
\begin{equation}
\left.\sqrt{\sigma_D}\right|_{T=T_c+} = (34.5 \pm 4.9)\,\mathrm{MeV}
\end{equation}
in the free energy, which should be compared to the ordinary zero-temperature
string tension $\sigma_0 = (440\,\mathrm{MeV})^2$ setting the overall scale.
This demonstrates again the weakness of the first order transition for
the colour group $G=SU(3)$. Quantitatively, our findings are in fair
agreement with lattice caluclations \cite{V9} which seem to favour
a slightly larger value $\sigma_D \simeq (48\,\mathrm{MeV})^2$.

\section{Further remarks and conclusions}

In this talk I have presented an effective model for the infra-red sector
of $SU(3)$ Yang Mills theory, based on random world surfaces carrying
center flux. Assuming only that vortex sheets have a surface tension and
stiffness, the model reproduces many non-trivial properties of long-range
Yang-Mills theory. Among the examples discussed here are the spatial
string tension, the order and strength of the phase transition for various
gauge groups, and the discontinuity of the vortex free energy across the
$SU(3)$ phase transition.

The geometrical structure of vortex branchings is the key property in
establishing first order behaviour for the $SU(3)$ model. Geometry alone,
however, is not sufficient to determine the physics of a vortex model; the
effective vortex dynamics specific to each gauge group play a crucial role.
In principle, the vortex dynamics would be determined by integrating out all
non-vortex degrees of freedom; our project is essentially the reverse approach
of guessing a model action and comparing to low-energy Yang-Mills theory.

For $SU(3)$, the results indicate that the simple two-operator action,
eq.~(\ref{eqn1}), is sufficient. With increasing complexity of the colour group,
however, additional terms may arise which favour vortex branchings (center monopoles)
\emph{explicitly}; such terms tend to enhance the strength of the phase transition.
In fact, preliminary investigations for $G=SU(4)$ indicate that such terms
are indeed necessary and may become even more pronounced with increasing number
of colours.



\end{document}